\documentclass[aps,prb,twocolumn,groupedaddress]{revtex4}

\usepackage{graphicx}

\begin{document}

\title{Identifying structural patterns in `disordered' metal clusters}

\author{Jonathan P.~K.~Doye}
\email[]{jpkd1@cam.ac.uk}
\affiliation{University Chemical Laboratory, Lensfield Road, Cambridge CB2 1EW, United Kingdom} 

\date{\today}

\begin{abstract}
Zinc and cadmium clusters interacting with a Gupta potential have previously
been identified as prototypical metallic systems that exhibiting disordered cluster structures.
Here, putative global minima of the potential energy have been located for these clusters for all sizes up
to $N\le 125$. Although none of the usual structural forms are lowest in energy and 
many of the clusters have no overall order, strong structural preferences have been identified.
Many of the clusters are based on distorted oblate Marks decahedra, where the distortion
involves the bringing together of 
atoms on either side of a re-entrant groove
of the Marks decahedron.
\end{abstract}

\maketitle

\section{\label{sect:intro}Introduction}
There has been much recent theoretical interest in the possibility that 
small clusters could have lowest-energy structures that are disordered
or amorphous. Examples have been found for model clusters interacting
with long-ranged pair potentials,\cite{Doye95c,Doye97d} and a wide 
variety of metal 
clusters.\cite{Glossman,Garzon98,Doye98c,Garzon99,Michaelian99,Garzon00,Wilson00,Taneda01,Michaelian02,Oviedo02,Massen02,Darby02,Wang02,Doye03a}
For some of these examples, these disordered clusters appear at sizes 
between the magic numbers for the usual icosahedral, decahedral or 
face-centred-cubic (fcc) forms, that are typically most stable for materials that
are close-packed for bulk;\cite{Doye98c,Wilson00,Darby02,Massen02} 
i.e.\ the disordered structures are more stable than
ordered structures with incomplete outer layers.
More interesting are the more extreme examples, where disordered structures
are lowest in energy even at the magic numbers of the common structural 
forms.\cite{Garzon98,Garzon99,Michaelian99,Michaelian02}
These results would suggest for these clusters that the disordered structures are 
dominant for the relevant size ranges.

By disordered it is usually meant that the structure has no discernible
overall order. Of course, there is still local order, as evidenced by 
structural probes, such as the radial distribution function, but these
usually have forms similar to that for bulk liquids and glasses.
Additionally, another common features for these clusters is that
there are many other disordered structures that lie very close in energy
to the global minimum.

Disordered clusters are most likely to occur when the constraints on the 
nearest-neighbour distances are weak. Then, the energetic cost for 
the strain present in the individual nearest-neighbour bonds can 
be low enough to be offset by other advantageous features of 
the disordered clusters, such as a low surface energy.
For the Morse clusters, this occurs when the potential is long-ranged
and has a wide, soft well.\cite{Doye95c,Doye97d} 
For metal clusters, the many-body part of the potential is
relatively insensitive to disorder in the nearest-neighbour 
distances.\cite{Soler00} 
If the contraction at the surface of the cluster (again due to 
the many-body forces) is strong, the ordered structures can be
sufficiently destabilized to make the disordered structures
lowest in energy.\cite{Soler00} 

For most of the examples where disordered structures have been found 
to be lower in energy than the most stable icosahedral, decahedral and 
fcc clusters, only a few sizes have usually been 
considered.\cite{Garzon98,Garzon99,Michaelian99,Michaelian02}
So, the possibility remains that at some other sizes the local 
structural preferences present in the disordered clusters 
could be assembled in a way that gives rise to a particularly stable 
structure with overall order. 
If this were the case, global optimization over a complete range 
of sizes would then reveal these new structural motifs and magic numbers.
This has recently been done for model lead clusters, where,
although none of the usual forms are ever lowest in energy and
many of the clusters have no apparent overall order, particularly
stable high-symmetry clusters were found at some sizes.\cite{Doye03a}  

Indeed there are general grounds for expecting high symmetry structures
to be more prevalent among the global minima, because such structures
are likely to have more extreme values of the energy (both high and low).\cite{Wales98}
This expectation, in my experience at least, seems to be born out empirically. 
It is rare that the global optimization of clusters does not reveal 
ordered high symmetry forms at some sizes. 
For example, even for potentials which have been designed to favour
glassy configurations, global optimization has revealed the presence of 
unusual, but nevertheless ordered, structures.\cite{Doye01a,Doye03b}

Here, I wish to examine some prototypical metal clusters that so far
have only been found to exhibit disordered structures. 
For potentials of the Gupta form the dependence of the tendency to disorder
on some of the parameters of the potential has been 
elaborated.\cite{Michaelian99,Soler00} 
Of the parameterized metals zinc and cadmium clusters emerged as those
with the strongest preference for disorder. This tendency was born out
in global optimization studies at sizes where particularly stable
fcc, decahedral and icosahedral structures were possible; as expected, 
the resulting structures appeared disordered.\cite{Michaelian02}

In this paper, I will attempt to identify structural patterns for these
two cluster systems, in particular searching for new magic numbers
and novel types of order.
This aim is in a similar spirit to Ref.\ \onlinecite{Soler01}, 
where possible structural patterns for gold clusters modelled
by the Gupta potential have been suggested.
To achieve this I have performed global optimization for all clusters
with up to 125 atoms. A further aim is to then relate back the identified
structural patterns to the form of the potential.

\section{\label{sect:methods}Methods}

\subsection{\label{sect:potential}Potential}

To model the zinc and cadmium clusters I use a Gupta  
potential\cite{Gupta} fitted by Cleri and Rosato.\cite{Cleri93} 
The potential energy is given by
\begin{eqnarray}
E&=&E_{\rm pair} + E_{\rm embed} \\
 &=&\sum_{i<j} \phi\left(r_{ij}\right)+\sum_{i} F\left(\bar\rho_i\right)\label{eq:EAM}, 
\end{eqnarray}
where $\phi(r)$ is a short-ranged pair potential, 
$U(\bar\rho)$ is a many-body embedding (or glue) function
and $\bar\rho_i$ is defined as 
\begin{equation}
\bar\rho_i=\sum_j \rho\left(r_{ij}\right),
\end{equation}
where $\rho(r)$ is an ``atomic density'' function.

\begin{table}
\caption{\label{table:params}Parameters for the Gupta potentials of Zn 
and Cd.}
\begin{ruledtabular}
\begin{tabular}{ccccc}
  & $p$ & $q$ & $A$ / eV & $\xi$ / eV \\
\hline
 Zn &  9.689 & 4.602 & 0.1477 & 0.8900 \\
 Cd & 10.612 & 5.206 & 0.1420 & 0.8117 \\
\end{tabular}
\end{ruledtabular}
\end{table}

For potentials of the Gupta form
\begin{eqnarray}
\phi\left(r\right)&=&2A e^{-p(r/r_0-1)}\\
F\left(\bar\rho\right)&=&-\xi \sqrt{\bar\rho} \\
\rho\left(r\right)&=&e^{-2q(r/r_0-1)}.
\end{eqnarray}
These forms arise from the second moment approximation
of a tight-binding Hamiltonian.
However, these functions are non-unique. 
Functions that give exactly the same energy can be 
constructed by the transformation
\begin{eqnarray}
\phi'\left(r\right)&=&\phi\left(r\right) + 2 g \rho(r) \\
F'\left(\bar\rho\right)&=&F\left(\bar\rho\right) - g \bar\rho .
\end{eqnarray}
This transformation redistributes the total energy between 
$E_{\rm pair}$ and $E_{\rm embed}$.
When 
\begin{equation}
g=\left.{dF\over d\bar\rho}\right|_{\bar\rho=\bar\rho_{\rm xtal}},
\end{equation}
$F'(\bar\rho)$ has a minimum at $\bar\rho_{\rm xtal}$,
where $\bar\rho_{\rm xtal}$ is the value of $\bar\rho$ in the equilibrium crystal.
This choice is called the effective pair format, and 
has been suggested as the most natural way to partition the 
energy between the pair and many-body contributions.\cite{Johnson89}
In this format, when $\bar\rho=\bar\rho_{\rm xtal}$ the 
pair potential controls the energy change for any change of configuration 
that does not significantly alter $\bar\rho$, hence the name. 
More specifically, by performing a Taylor expansion about this reference
density, one can show that to first order, the change in energy is 
due to the pair potential.
Consequently, it is also a much more helpful format for relating 
the structure to the form of the potential.

The Gupta potential in this effective pair format becomes
\begin{eqnarray}
\phi_{\rm eff}\left(r\right)&=&2Ae^{-p(r/r_0-1)} -  
    {\xi\over\sqrt{\bar\rho_{\rm xtal}}} e^{-2q(r/r_0-1)}\\
F_{\rm eff}\left(\bar\rho\right)&=&-\xi \sqrt{\bar\rho}
 \left(1-{1\over 2}\sqrt{\bar\rho\over\bar\rho_{\rm xtal}}\right).
\end{eqnarray}
$\phi_{\rm eff}(r)$ is a sum of two exponentials, so 
for $p>2q$ it is repulsive at short range and has an attractive well.
The minimum in $\phi_{\rm eff}(r)$ is at 
\begin{equation}
r_{\rm min}=r_0\left(1+{1\over p-2q}\log\left(
  {Ap\sqrt{\bar\rho_{\rm xtal}}\over\xi q}\right)\right)
\end{equation}
and is of depth
\begin{equation}
\phi_{\rm eff}\left(r_{\rm min}\right)=-A\left(
    {\xi q\over Ap\sqrt{\bar\rho_{\rm xtal}}}\right)^{p/p-2q}
    \left({p-2q\over q}\right)
\end{equation}
As $p\rightarrow 2q$ from above, the depth of the well in the 
pair potential goes to zero.
$F_{\rm eff}(\bar\rho)$ is quadratic in $\sqrt{\bar\rho}$ and
has a minimum of depth $-\xi\sqrt{\bar\rho_{\rm xtal}}/2$ at 
$\bar\rho=\bar\rho_{\rm xtal}$.

\begin{figure}
\includegraphics[width=8.4cm]{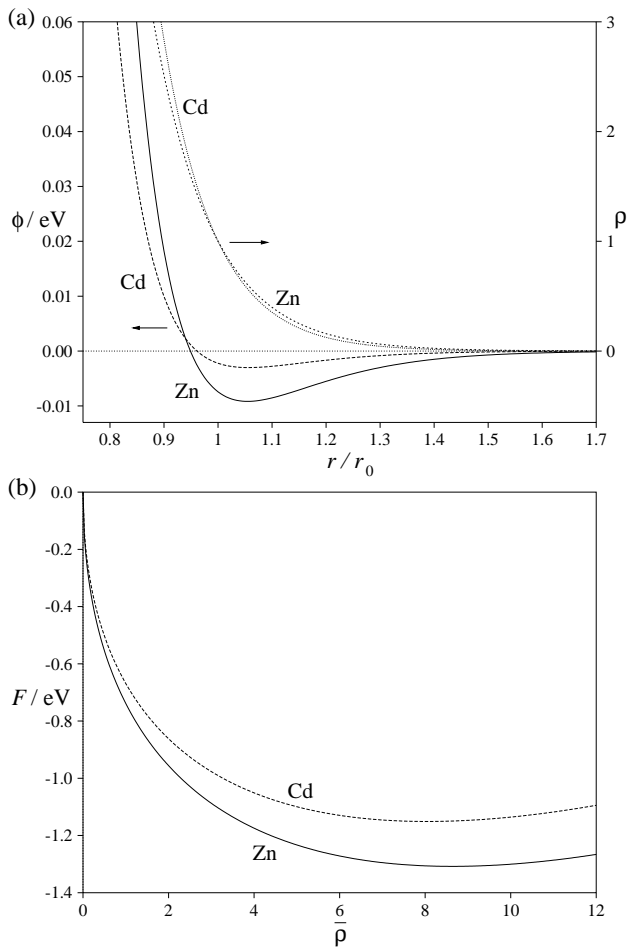}
\caption{\label{fig:potential}The three functions that make up the 
potential:
(a) $\phi_{\rm eff}(r)$, $\rho(r)$ and (b) $F_{\rm eff}(\bar\rho)$.}
\end{figure}

The Gupta parameters for the Zn and Cd potentials are given in
Table \ref{table:params}, and the functions $\phi_{\rm eff}$, $\rho$ and 
$F_{\rm eff}$ are shown in Fig.\ \ref{fig:potential}. 
The shallowness of the attractive well in the 
effective pair potential of cadmium is particularly apparent---it is only
0.26\% of the depth of the minimum in the embedding function.
The pair well depth is somewhat larger for zinc, but it
is still only 0.70\% of the embedding function minimum.
For both systems the shallowness arises because $p/q$ is close to 2.

This feature of the potentials will have similar structural consequences
for both systems. Firstly, the majority of the binding energy will come
from the many-body interactions. Secondly, the pair potential provides 
relatively little constraint on the pair distances, except for the repulsion
at short distances. Therefore, the most important feature for obtaining
a low-energy configuration is to have the individual $\bar\rho_i$ 
values as close as possible to the optimal value. 
By contrast, there is little energetic advantage in having the 
nearest-neighbour distances close to the minimum of the pair potential.
For a cluster the former can sometimes be more easily achieved through a 
disordered structure than one based on a lattice, because the additional
flexibility of not having well-defined nearest-neighbour distances
makes it easier to obtain close to optimal $\bar\rho$ values for the
surface atoms. This is the source of the tendency to disorder for zinc and
cadmium clusters described by the current potentials.

For bulk both zinc and cadmium are hexagonal close-packed with
$\bar\rho_{\rm xtal}^{\rm Cd}=8.042$, and
$\bar\rho_{\rm xtal}^{\rm Zn}=8.638$. These are anomalously low values
for close-packed materials. For example, for an ideal close-packed crystal 
with all nearest-neighbours at $r_0$, the nearest-neighbour contribution to $\bar\rho$
is 12. However, the ratios of the unit cell parameters $c/a$ for the Zn
and Cd crystals are particularly large, 
and so the contribution to $\bar\rho$ from 
nearest-neighbour distances with a component in the $c$ direction is 
significantly reduced. 

The low values of $\bar\rho_{\rm xtal}$ have structural consequences
for the clusters.
An atom can achieve the optimal $\bar\rho$ values with only eight or nine 
neighbours, so this is possible for surface atoms. 
By contrast, the nearest-neighbour distances for atoms in the 
interior need to be elongated to prevent unfavourably large values of $\bar\rho$. 
This is bad news for cluster structures, such as the
Mackay icosahedron, where the interior distances are naturally shorter than
those on the surface.

The exponential nature of $\rho(r)$ means that significant changes to 
the $\bar\rho$ values can be achieved by relatively small changes to the 
nearest-neighbour distances. Consequently, it is somewhat easier for
atoms, even those with low coordination numbers, to obtain nearly optimal
$\bar\rho$ values. This behaviour contrasts with other many-body potential,
for instance those produced by the force-matching method,\cite{Ercolessi94} where there is no
presumed form for $\rho(r)$, and so more long-ranged functions can 
result.\cite{Lim,Ercolessi94,Mishin99} 

The current potentials were obtained by keeping the $c/a$ value fixed at the
experimental value. This is at some cost in terms of the quality of the
fit to other quantities. This led Cleri and Rosato to also construct potentials
in which the $c/a$ ratio was allowed to vary in the fitting procedure,
and resulted in a better quality fit for many properties.\cite{Cleri93} 
Such a potential is available for cadmium and has dramatically different 
structural properties because $p/q=3.49$.

\subsection{\label{sect:gopt}Global Optimization}

The global optimization of the zinc and cadmium clusters 
was performed using the basin-hopping \cite{WalesD97,WalesS99}
(or Monte Carlo minimization \cite{Li87a}) approach. 
This method has proved particularly
successful in locating putative global minima for a 
wide variety of cluster systems.
The nature of the potentials for the 
current clusters, particularly that $p-2q$ is small,
makes global optimization difficult compared to many other metal 
clusters of similar size.
Therefore, a considerable computational effort was required
to extend the results up to 125 atoms. 
It proved particularly important to supplement the
standard unbiased runs from random starting points for each size, 
with short runs started from low-energy minima of nearby sizes with the 
appropriate number of atoms added or removed. 
As the structures of the two clusters are similar---seventy of the 
global minima are the same---it also proved useful to reoptimize the 
low-energy minima found for one metal for the other.
These two types of `seeded' runs were applied iteratively until no further 
new global minima were located.

It should be noted that, of course, 
there is no guarantee that I have been able to locate the true global 
minima, and the probability that a global minimum has been missed
will increase with cluster size, 
as the size of the search space, and hence the number of 
minima,\cite{Tsai93a,Still99,Doye02a} increases exponentially with $N$.
Examination of the statistics of how often independent runs 
locate the same lowest-energy minimum and the importance of the
seeded runs can provide one with an idea of the likelihood that 
the true global minimum has been found.
For example, for zinc clusters with less than ninety atoms 
virtually all the putative global minima were located in unbiased runs, 
but for $N>100$ the majority of putative global minima were only located 
in seeded runs. Similarly, for the cadmium clusters virtually
all the putative global minima were only located in seeded runs 
for $N\ge65$. This greater difficulty is because $2p-q$ is closer to zero
for cadmium.
However, given the similarity of the observed structures for the two metals, 
I am confident that the vast majority of the putative global minima up 
to $N=100$ cannot be bettered, but beyond this size one's degree of 
scepticism about the success of the global optimization should increase 
rapidly.

\begin{table*}
\caption{\label{table:gmin_Zn}Energies (in eV) and point groups of putative Zn$_N$ global minima}
\begin{ruledtabular}
\begin{tabular}{ccccccccccccccccccc}
$N$ & PG & Energy & & $N$ & PG & Energy & & $N$ & PG & Energy & & $N$ & PG & Energy & & $N$ & PG & Energy \\
\hline
   3 &  $D_{3h}$ &    -3.751296 & &   28 &  $D_{6d}$ &   -37.076416 & &   53 &  $C_1$    &   -70.622509 & &   78 &  $C_1$    &  -104.238674 & &  103 &  $C_1$    &  -137.902829 \\ 
   4 &  $T_d$    &    -5.099346 & &   29 &  $C_2$    &   -38.410947 & &   54 &  $C_1$    &   -71.969642 & &   79 &  $C_{2v}$ &  -105.595357 & &  104 &  $C_1$    &  -139.255744 \\ 
   5 &  $D_{3h}$ &    -6.422287 & &   30 &  $C_1$    &   -39.768212 & &   55 &  $C_1$    &   -73.310032 & &   80 &  $C_{2v}$ &  -106.927940 & &  105 &  $C_1$    &  -140.593303 \\ 
   6 &  $O_h$    &    -7.765873 & &   31 &  $C_1$    &   -41.107234 & &   56 &  $C_1$    &   -74.649595 & &   81 &  $C_s$    &  -108.284296 & &  106 &  $C_1$    &  -141.928489 \\ 
   7 &  $D_{5h}$ &    -9.088519 & &   32 &  $C_1$    &   -42.444864 & &   57 &  $C_1$    &   -75.992628 & &   82 &  $C_s$    &  -109.625208 & &  107 &  $C_1$    &  -143.279843 \\ 
   8 &  $D_{2d}$ &   -10.409083 & &   33 &  $C_1$    &   -43.776617 & &   58 &  $C_1$    &   -77.336947 & &   83 &  $C_s$    &  -110.976689 & &  108 &  $C_1$    &  -144.635048 \\ 
   9 &  $C_{2v}$ &   -11.733002 & &   34 &  $C_3$    &   -45.125218 & &   59 &  $C_1$    &   -78.672227 & &   84 &  $C_s$    &  -112.322650 & &  109 &  $C_1$    &  -145.978993 \\ 
  10 &  $C_{3v}$ &   -13.055156 & &   35 &  $C_s$    &   -46.472931 & &   60 &  $C_1$    &   -80.019432 & &   85 &  $C_1$    &  -113.668080 & &  110 &  $C_1$    &  -147.337545 \\ 
  11 &  $C_2$    &   -14.379679 & &   36 &  $D_{2h}$ &   -47.814946 & &   61 &  $C_2$    &   -81.363067 & &   86 &  $C_{2v}$ &  -115.021815 & &  111 &  $C_2$    &  -148.697585 \\ 
  12 &  $C_2$    &   -15.706971 & &   37 &  $C_1$    &   -49.144718 & &   62 &  $C_1$    &   -82.708058 & &   87 &  $C_{2v}$ &  -116.375101 & &  112 &  $C_1$    &  -150.033337 \\ 
  13 &  $C_s$    &   -17.032142 & &   38 &  $C_s$    &   -50.499983 & &   63 &  $C_1$    &   -84.061400 & &   88 &  $C_{2v}$ &  -117.726908 & &  113 &  $C_1$    &  -151.385243 \\ 
  14 &  $C_{6v}$ &   -18.369312 & &   39 &  $C_{2v}$ &   -51.846864 & &   64 &  $C_2$    &   -85.419385 & &   89 &  $C_s$    &  -119.057698 & &  114 &  $C_1$    &  -152.720206 \\ 
  15 &  $C_s$    &   -19.697479 & &   40 &  $C_s$    &   -53.182148 & &   65 &  $C_1$    &   -86.749702 & &   90 &  $C_s$    &  -120.405576 & &  115 &  $C_2$    &  -154.071651 \\ 
  16 &  $C_s$    &   -21.042019 & &   41 &  $C_s$    &   -54.521834 & &   66 &  $C_1$    &   -88.097872 & &   91 &  $C_{2v}$ &  -121.767147 & &  116 &  $C_1$    &  -155.420956 \\ 
  17 &  $C_{2v}$ &   -22.395676 & &   42 &  $D_4$    &   -55.874358 & &   67 &  $C_1$    &   -89.426800 & &   92 &  $C_{2v}$ &  -123.118340 & &  117 &  $C_1$    &  -156.762711 \\ 
  18 &  $C_{4v}$ &   -23.735972 & &   43 &  $C_4$    &   -57.213737 & &   68 &  $C_1$    &   -90.782264 & &   93 &  $C_s$    &  -124.451825 & &  118 &  $C_1$    &  -158.109816 \\ 
  19 &  $D_{4d}$ &   -25.067563 & &   44 &  $C_s$    &   -58.554345 & &   69 &  $C_1$    &   -92.135925 & &   94 &  $C_s$    &  -125.796431 & &  119 &  $C_1$    &  -159.451202 \\ 
  20 &  $C_{2v}$ &   -26.395373 & &   45 &  $C_{2v}$ &   -59.895318 & &   70 &  $C_1$    &   -93.473840 & &   95 &  $C_s$    &  -127.142167 & &  120 &  $C_{2v}$ &  -160.812498 \\ 
  21 &  $C_s$    &   -27.715028 & &   46 &  $C_s$    &   -61.227348 & &   71 &  $C_s$    &   -94.828826 & &   96 &  $C_{2v}$ &  -128.501419 & &  121 &  $C_1$    &  -162.152965 \\ 
  22 &  $C_{2v}$ &   -29.042499 & &   47 &  $C_s$    &   -62.580198 & &   72 &  $C_s$    &   -96.168254 & &   97 &  $C_s$    &  -129.836679 & &  122 &  $C_1$    &  -163.502185 \\ 
  23 &  $C_1$    &   -30.386460 & &   48 &  $C_s$    &   -63.916581 & &   73 &  $C_s$    &   -97.513085 & &   98 &  $C_{2v}$ &  -131.170669 & &  123 &  $C_{2v}$ &  -164.859277 \\ 
  24 &  $C_2$    &   -31.728122 & &   49 &  $C_{2v}$ &   -65.269233 & &   74 &  $C_s$    &   -98.853247 & &   99 &  $C_s$    &  -132.525475 & &  124 &  $C_1$    &  -166.196591 \\ 
  25 &  $C_2$    &   -33.075590 & &   50 &  $C_s$    &   -66.601903 & &   75 &  $C_{2v}$ &  -100.203580 & &  100 &  $C_{2v}$ &  -133.883827 & &  125 &  $C_1$    &  -167.548063 \\ 
  26 &  $C_2$    &   -34.412179 & &   51 &  $C_1$    &   -67.938575 & &   76 &  $C_1$    &  -101.545341 & &  101 &  $C_s$    &  -135.219057 \\ 
  27 &  $C_2$    &   -35.744950 & &   52 &  $C_{2v}$ &   -69.281361 & &   77 &  $C_s$    &  -102.900278 & &  102 &  $C_1$    &  -136.566429 \\ 
\end{tabular}
\end{ruledtabular}
\end{table*}

\section{\label{sect:gmin}Global minima}

\begin{figure*}
\includegraphics[width=18cm]{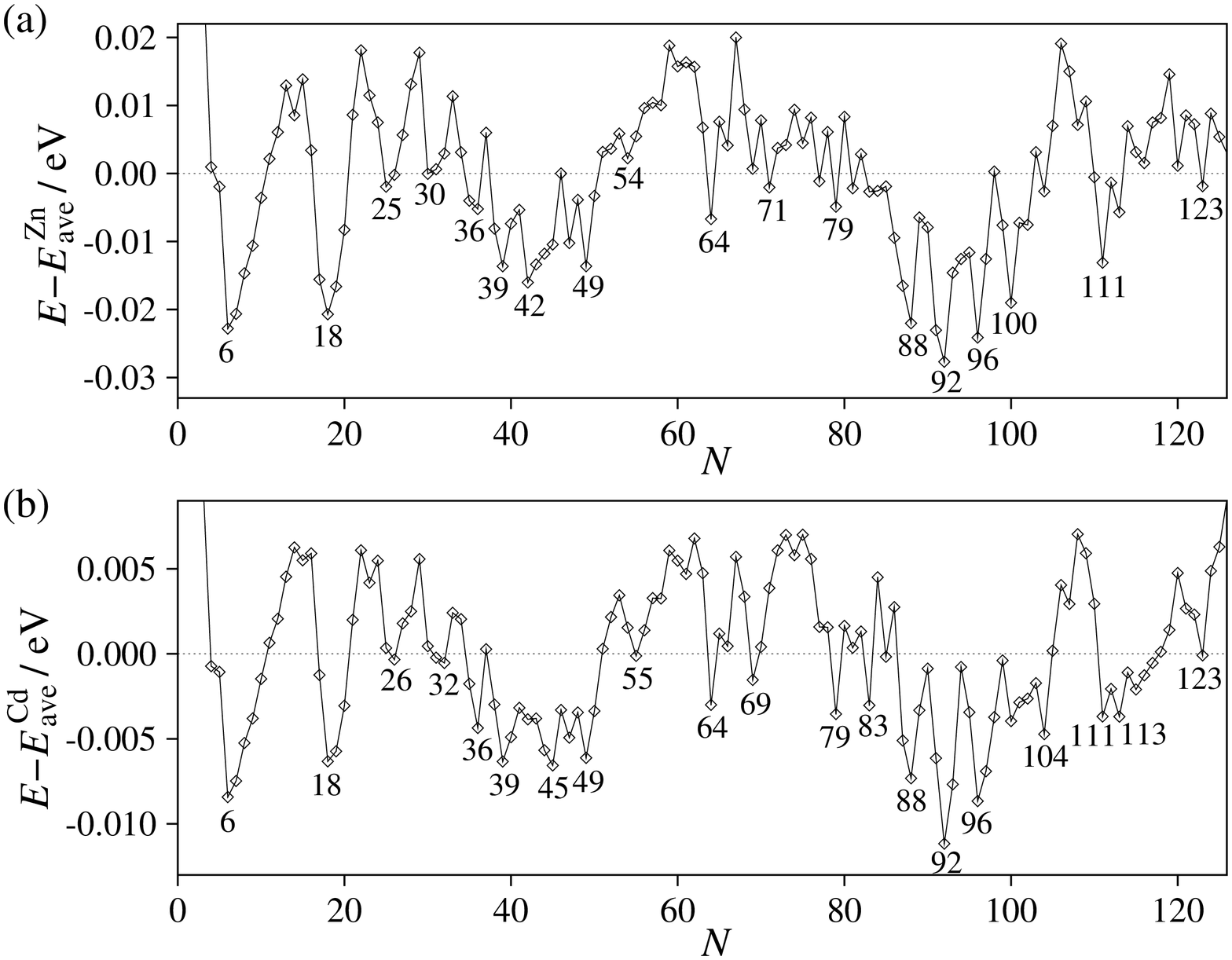}
\caption{\label{fig:EvN}Energies of the putative global minima of (a) Zn$_N$ and (b) Cd$_N$ relative 
to $E_{\rm ave}$, a four-parameter fit to their energies.
$E_{\rm ave}^{\rm Zn}=-1.3621 N + 0.1062 N^{2/3} - 0.0067 N^{1/3} + 0.0913$.
$E_{\rm ave}^{\rm Cd}=-1.1684 N + 0.0391 N^{2/3} - 0.0143 N^{1/3} + 0.0425$.}
\end{figure*}

The energies and point groups for the putative global minima are 
given in Tables \ref{table:gmin_Zn} and \ref{table:gmin_Cd}.
Point files are available online at the Cambridge Cluster Database.\cite{Web}
The energies of the global minima are represented in Fig.\ \ref{fig:EvN} 
in such a way that makes particularly stable clusters stand out. 

\begin{table*}
\caption{\label{table:gmin_Cd}Energies (in eV) and point groups of putative Cd$_N$ global minima. 
Those labelled with a star have the same structure as the Zn$_N$ global minimum.}
\begin{ruledtabular}
\begin{tabular}{cccccccccccccccccccccccc}
$N$ & PG & Energy & & & $N$ & PG & Energy & & & $N$ & PG & Energy & & & $N$ & PG & Energy & & & $N$ & PG & Energy  & \\
\hline
  3 &  $D_{3h}$ &   -3.391649 & * & &  28 &  $C_2$    &  -32.354453 &   & &  53 &  $C_1$    &  -61.383718 &   & &   78 &  $C_1$    &   -90.441924 & * & &  103 &  $C_1$    &  -119.516811 & * \\ 
  4 &  $T_d$    &   -4.556244 & * & &  29 &  $C_1$    &  -33.511796 &   & &  54 &  $C_1$    &  -62.547483 &   & &   79 &  $C_{2v}$ &   -91.609644 & * & &  104 &  $C_1$    &  -120.682922 & * \\ 
  5 &  $D_{3h}$ &   -5.710992 & * & &  30 &  $C_2$    &  -34.677431 &   & &  55 &  $C_s$    &  -63.711040 &   & &   80 &  $C_s$    &   -92.767096 &   & &  105 &  $C_1$    &  -121.841138 &   \\ 
  6 &  $O_h$    &   -6.873544 & * & &  31 &  $C_1$    &  -35.838691 & * & &  56 &  $C_2$    &  -64.871474 &   & &   81 &  $C_s$    &   -93.931045 & * & &  106 &  $C_1$    &  -123.000416 & * \\ 
  7 &  $D_{5h}$ &   -8.028455 & * & &  32 &  $C_1$    &  -36.999668 & * & &  57 &  $C_1$    &  -66.031560 & * & &   82 &  $C_1$    &   -95.092765 &   & &  107 &  $C_1$    &  -124.164675 & * \\ 
  8 &  $D_{2d}$ &   -9.182587 & * & &  33 &  $C_1$    &  -38.157466 & * & &  58 &  $C_1$    &  -67.193590 &   & &   83 &  $C_s$    &   -96.259818 & * & &  108 &  $C_1$    &  -125.323742 &   \\ 
  9 &  $C_{2v}$ &  -10.337961 & * & &  34 &  $C_1$    &  -39.318659 &   & &  59 &  $C_1$    &  -68.352805 & * & &   84 &  $C_1$    &   -97.415006 &   & &  109 &  $C_1$    &  -126.488046 &   \\ 
 10 &  $D_{4d}$ &  -11.492834 &   & &  35 &  $C_1$    &  -40.483344 &   & &  60 &  $C_1$    &  -69.515486 &   & &   85 &  $C_1$    &   -98.582433 & * & &  110 &  $C_1$    &  -127.654226 & * \\ 
 11 &  $C_2$    &  -12.648250 & * & &  36 &  $D_{2h}$ &  -41.646892 & * & &  61 &  $C_2$    &  -70.678373 & * & &   86 &  $C_1$    &   -99.742283 &   & &  111 &  $C_2$    &  -128.824065 & * \\ 
 12 &  $C_2$    &  -13.804680 & * & &  37 &  $C_s$    &  -42.803277 &   & &  62 &  $C_2$    &  -71.838429 &   & &   87 &  $C_{2v}$ &  -100.912912 & * & &  112 &  $C_1$    &  -129.985671 & * \\ 
 13 &  $C_s$    &  -14.960307 & * & &  38 &  $D_{2d}$ &  -43.967614 &   & &  63 &  $C_1$    &  -73.002649 & * & &   88 &  $C_{2v}$ &  -102.077956 & * & &  113 &  $C_1$    &  -131.150549 & * \\ 
 14 &  $D_{6d}$ &  -16.116921 &   & &  39 &  $C_{2v}$ &  -45.132103 & * & &  64 &  $C_2$    &  -74.172601 & * & &   89 &  $C_s$    &  -103.236780 & * & &  114 &  $C_1$    &  -132.311215 & * \\ 
 15 &  $D_3$    &  -17.276238 &   & &  40 &  $C_s$    &  -46.291870 & * & &  65 &  $C_1$    &  -75.330653 &   & &   90 &  $C_1$    &  -104.397182 &   & &  115 &  $C_2$    &  -133.475490 & * \\ 
 16 &  $C_s$    &  -18.434583 &   & &  41 &  $C_{2v}$ &  -47.451405 &   & &  66 &  $C_1$    &  -76.493660 & * & &   91 &  $C_{2v}$ &  -105.565322 & * & &  116 &  $C_1$    &  -134.637941 & * \\ 
 17 &  $C_{2v}$ &  -19.600675 & * & &  42 &  $D_4$    &  -48.613389 & * & &  67 &  $C_1$    &  -77.650700 &   & &   92 &  $C_{2v}$ &  -106.733237 & * & &  117 &  $C_1$    &  -135.800525 & * \\ 
 18 &  $C_{2v}$ &  -20.764865 &   & &  43 &  $C_4$    &  -49.774701 & * & &  68 &  $C_1$    &  -78.815370 & * & &   93 &  $C_s$    &  -107.892659 & * & &  118 &  $C_1$    &  -136.963187 &   \\ 
 19 &  $C_{2v}$ &  -21.923540 &   & &  44 &  $C_s$    &  -50.937992 & * & &  69 &  $C_1$    &  -79.982620 & * & &   94 &  $C_s$    &  -109.048698 & * & &  119 &  $C_1$    &  -138.125231 &   \\ 
 20 &  $C_{2v}$ &  -23.080283 & * & &  45 &  $C_{2v}$ &  -52.100346 & * & &  70 &  $C_1$    &  -81.143066 & * & &   95 &  $C_s$    &  -110.214298 &   & &  120 &  $C_1$    &  -139.285228 &   \\ 
 21 &  $C_1$    &  -24.234793 &   & &  46 &  $C_s$    &  -53.258607 &   & &  71 &  $C_1$    &  -82.302019 &   & &   96 &  $C_s$    &  -111.382493 &   & &  121 &  $C_2$    &  -140.450672 &   \\ 
 22 &  $C_1$    &  -25.390380 &   & &  47 &  $C_s$    &  -54.421794 & * & &  72 &  $C_1$    &  -83.462242 &   & &   97 &  $C_s$    &  -112.543725 & * & &  122 &  $C_1$    &  -141.614397 &   \\ 
 23 &  $C_1$    &  -26.552108 & * & &  48 &  $C_s$    &  -55.581917 & * & &  73 &  $C_s$    &  -84.623790 & * & &   98 &  $C_s$    &  -113.703547 &   & &  123 &  $C_{2v}$ &  -142.780180 & * \\ 
 24 &  $C_1$    &  -27.710728 &   & &  49 &  $C_{2v}$ &  -56.746239 & * & &  74 &  $C_s$    &  -85.787478 & * & &   99 &  $C_s$    &  -114.863226 &   & &  124 &  $C_1$    &  -143.938617 &   \\ 
 25 &  $C_2$    &  -28.875894 & * & &  50 &  $C_s$    &  -57.905181 & * & &  75 &  $C_1$    &  -86.948788 &   & &  100 &  $C_s$    &  -116.029828 &   & &  125 &  $C_1$    &  -145.100610 & * \\ 
 26 &  $C_2$    &  -30.036716 &   & &  51 &  $C_1$    &  -59.063265 &   & &  76 &  $C_1$    &  -88.112746 & * & &  101 &  $C_s$    &  -117.191782 & * \\ 
 27 &  $C_1$    &  -31.194825 &   & &  52 &  $C_1$    &  -60.223182 &   & &  77 &  $C_s$    &  -89.279311 & * & &  102 &  $C_1$    &  -118.354639 & * \\ 
\end{tabular}
\end{ruledtabular}
\end{table*}

The previous results for these clusters were for a selection of sizes 
that often show highly symmetric structures, namely $N=13$, 38, 55 and 
147.\cite{Michaelian02} 
For $N$=13 and 38 the three lowest-energy minima reported
in Ref.\ \onlinecite{Michaelian02} agree with the current results.
However, for Zn$_{55}$, Zn$_{75}$ and Cd$_{75}$ 
the lowest-energy structures reported by Michaelian {\it et al.} 
correspond at best to the third, eighth, and twenty-second 
lowest-energy minima, respectively, and lie 0.0031, 0.0186 and 0.0086$\,$eV
above the lowest-energy minima reported here. These energies are 
significant compared to the variations shown in Fig.\ \ref{fig:EvN}.
I did not systematically attempt to optimize 147-atom clusters, as locating
the true global minimum for this size would be extremely difficult. 
However, short basin-hopping runs did find structures that were 
0.0966 and 0.0304$\,$eV lower in energy for Zn$_{147}$ and Cd$_{147}$, 
respectively, than the lowest-energy structures found by 
Michaelian {\it et al.}\cite{Michaelian02}

Before, we examine the observed structures, there are a number of interesting
features evident from Fig.\ \ref{fig:EvN}. The energy zero in these
figures is $E_{\rm ave}$, a four-parameter fit to the energies of the global 
minima, where the first two terms correspond to volume and surface energies.
For these two clusters the surface term is exceptionally small. The ratio
of the surface to the volume coefficient is 7.8\% for Zn and 3.3\% for Cd.
For comparison the value of this ratio is 197\% for Lennard-Jones 
clusters,\cite{Doye99f} 48\% for Gupta lead clusters,\cite{Lai02,Doyeunpub} 
60\% for aluminium clusters,\cite{Doye03d} and 93\%, 87\% and 63\% for 
Sutton-Chen silver, nickel and gold clusters, respectively\cite{Doye98c}. 
One expects the surface
energy to be lower for metal clusters than for a cluster interacting
with a pair potential, because the lower coordinate surface atoms of a metal can
increase their many-body embedding energy by shortening their nearest neighbour
distances. But the Zn and Cd surface energies are very low even for metals,
because the optimal value of $\bar\rho$ is small enough to be
achievable by the surface atoms and the effective pair potential only 
makes a small contribution to the energy.

The second interesting feature is the magnitude of the fluctuations about
the average energy. For Cd and Zn they are again exceptionally small.
The average deviation from $E_{\rm ave}$ compared to the average energy 
per atom is 0.67\% for Zn and 0.31\% for Cd.
For comparison, the value is 16\% for Lennard-Jones clusters, 
2.1\% for Gupta lead clusters\cite{Lai02,Doyeunpub} and  5.6\% 
for aluminium clusters\cite{Doye03d} in the same size range.

The implications of these small fluctuations are that the differences 
between the more and less stable sizes indicated by Fig. \ref{fig:EvN}
are small, making it more difficult to observe any such 
`magic' numbers.  Interestingly, one would expect that the properties of 
fully disordered clusters would evolve smoothly with size.
By this measure, these zinc and cadmium clusters seem to be close to this limit.

\begin{figure*}
\includegraphics[width=18cm]{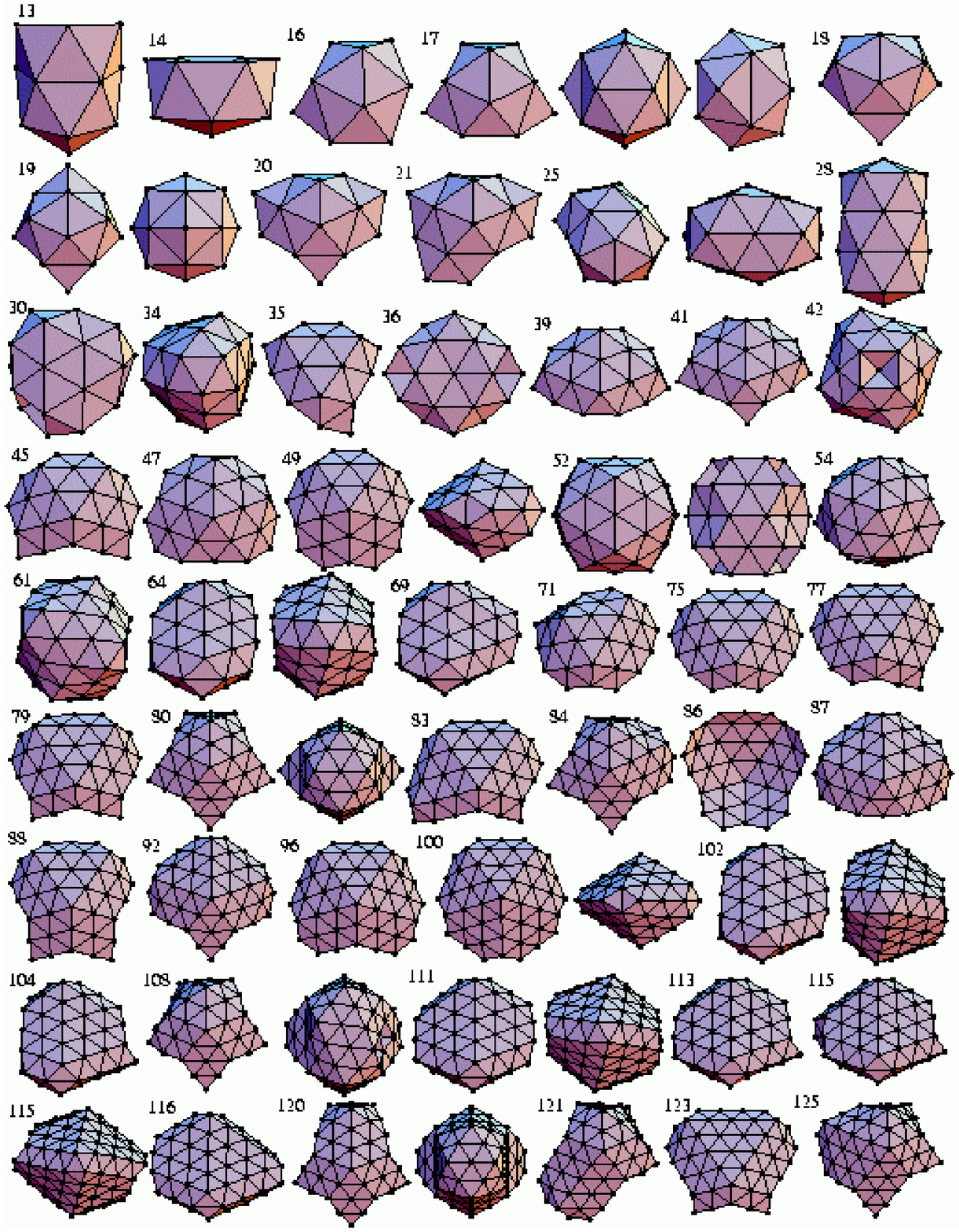}
\caption{\label{fig:Znselect}A selection of the putative global minima for Zn$_N$.}
\end{figure*}

We will first examine the zinc clusters in detail, and then later look at the 
relatively small structural difference between the two systems.
A selection of zinc clusters are depicted in Fig.\ \ref{fig:Znselect}.
that are either particularly stable or have some interesting structural feature. 
Up to $N=10$ the structure of the smallest zinc clusters are typical of 
what one usually finds for clusters modelled by empirical potentials. 
However, instead of structures leading up to the 13-atom icosahedra, more
open structures are then preferred. The example shown in Fig.\ \ref{fig:Znselect}
for Zn$_{13}$ can be considered as a polytetrahedral fragment of a 19-atom
double icosahedron. It is insightful to examine, why this structure is 
lower in energy than the 13-atom icosahedron.
The icosahedron is in fact $0.335\,$eV higher in energy, and this stems from
an unfavourable embedding energy. The better pair energy, because of the 
icosahedron's greater average coordination number, does little to offset this.
As the distances between the central atom 
and the vertices of the icosahedron are 5\% shorter than the distance 
between adjacent vertices and $\rho(r)$ increases rapidly
with decreasing $r$, 
the central atom has an extremely high value of $\bar\rho$,
namely 19.568, that is much greater than the optimal value. 
By contrast, the $\bar\rho$ values for all the atoms in the global minimum
are relatively close to $\rho_{\rm xtal}$. This is achieved by the
atoms with low coordination number having shorter average distances.

This example illustrates two important features of the potential. First,
low-coordinate surface atoms are not disfavoured, in contrast to a potential 
with a strong pair component. Second, it is important that interior atoms
do not have short distances. As we will see, it is often much better for 
an interior atom to have a coordination number larger than twelve since
then its nearest-neighbour distances will be longer than those between  
its neighbours.  
This latter feature is illustrated by the 14-atom global minimum. 
The 13-coordinate atom has an average nearest-neighbour separation of 
$1.052\,r_0$, whereas the average separation for the nearest-neighbour contacts 
not involving this atom is $0.958\,r_0$. Consequently $\bar\rho$ for the central 
atom is only 10.381

\begin{figure}
\includegraphics[width=8.4cm]{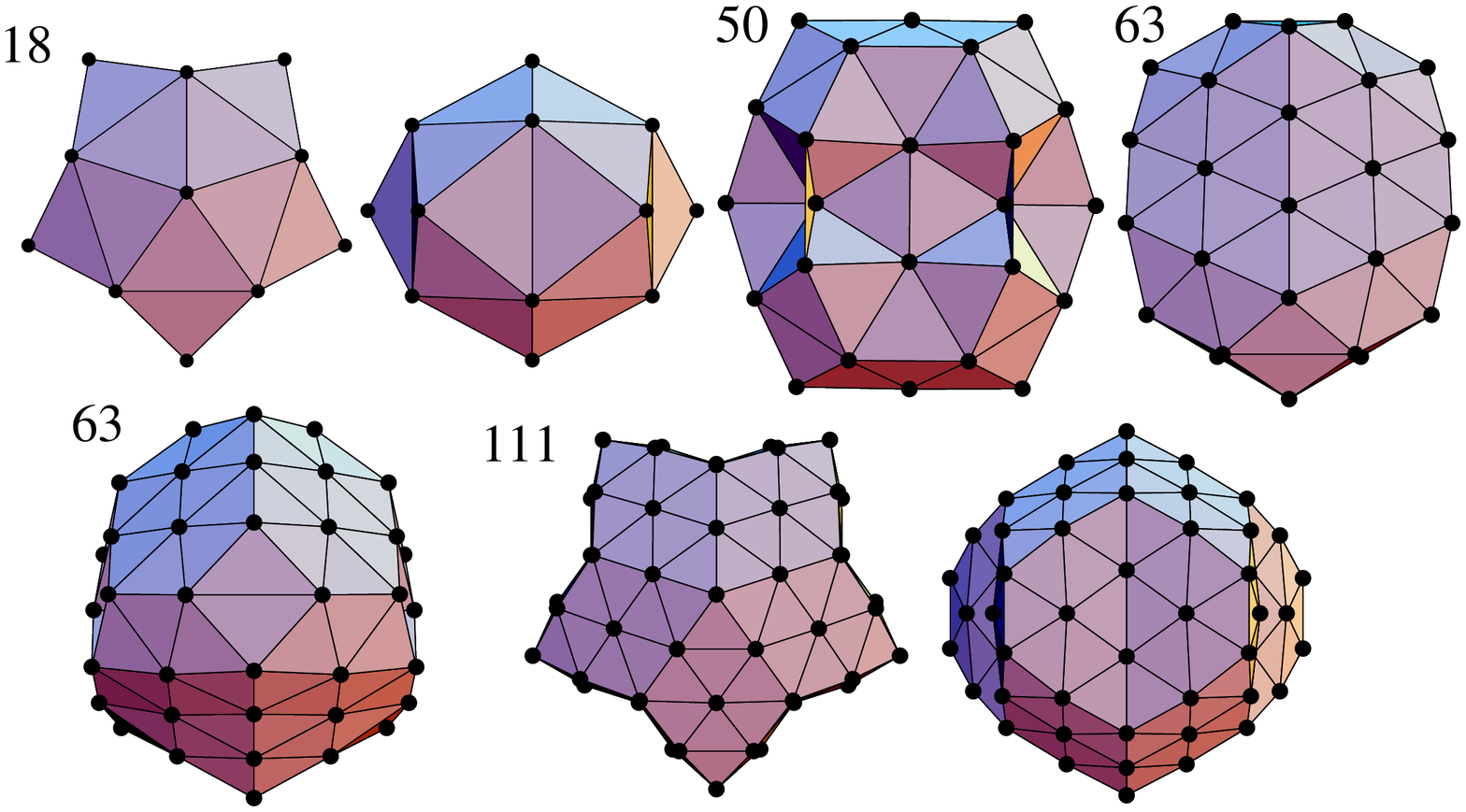}
\caption{\label{fig:ref_struct}A selection of reference
structures for interpreting the Zn$_N$ and Cd$_N$ global minima.}
\end{figure}

The global minima for $N$=16--21 can all be characterized as distorted
decahedral structures. Decahedral structures are based on pentagonal 
bipyramids (hence the name) and have a single five-fold axis of symmetry. 
The pentagonal bipyramids themselves are usually not particularly stable 
because they have a relatively non-spherical shape. By the introduction of 
reentrant grooves at the five equatorial vertices that are parallel to the 
five-fold axis, more stable Marks decahedra can be produced.\cite{Marks84,alsotruncate}
An 18-atom example is shown in Fig.\ \ref{fig:ref_struct} that 
was derived from a 23-atom pentagonal bipyramid, where 
the introduction of the grooves gives rise to five four-coordinate 
capping atoms.

This 18-atom Marks decahedron is in fact itself not that unstable
and is only 0.042$\,$eV above the global minimum. Even though
the central atom is 12-coordinate, this atom is able to maintain a reasonable
$\bar\rho$ value (9.867), whilst an expansion of the structure along 
the axial direction allows the surface distances to contract. 

However, the energy can be further improved by distorting this Marks 
decahedron. The distortion of this decahedra is easy to see for 
the 17-atom global minimum. This structure can be derived from the 
Marks decahedron (minus one capping one atom) by 
a diamond-square-diamond rearrangement,\cite{Lipscomb} where the contact 
between the atoms in the groove opposite the missing cap atom is broken 
and a contact between the adjacent capping atoms is formed (compare 
the second views of the two structures in Fig.\ \ref{fig:Znselect} 
and \ref{fig:ref_struct}). 
During this process the two capping atoms are also drawn closer to 
the central atom to form new contacts, thus giving the central atom a 
coordination number of 14. The planes containing the two five-fold rings
of atoms that were parallel in the Marks decahedron are now splayed apart
(see the third view of the Zn$_{17}$ global minimum).
It is also noticeable that the structure in the region of the distortion 
locally resembles the 13-atom icosahedron with the surface atoms now
six coordinate.\cite{Soler01} The second view of Zn$_{17}$ looks down
a two-fold axis of this icosahedral-like region.

The energetic advantage provided through this `groove-bridging' distortion
is mainly through the lowering of the energies of the two capping atoms
that come into contact. The increase in their coordination number to 6 gives
rise to both an increase in $\bar\rho$ and a more favourable pair energy.
The energy of the central atom is left relatively unaffected, because the
increase in coordination number is compensated by an increase in the average
nearest-neighbour separation for that atom.

The other global minima for $N$=16--21 can be also understood in terms of this
distortion. Zn$_{16}$ simply has one less capping atom than Zn$_{17}$. For $N$=18 and 19 
two of the grooves are bridged, giving rise to a 16-coordinate central atom.
The combined effect of the two distortions also causes the contact in the 
groove adjacent to the these two distortions to be broken, thus giving 
rise to a square face, which is itself capped in Zn$_{19}$, and 
a four-fold axis of symmetry (see the second view of Zn$_{19}$).
Zn$_{20}$ and Zn$_{21}$, like Zn$_{17}$, are only distorted at one groove;
the additional atoms fill in some of the other grooves 
taking the structure closer towards the 23-atom pentagonal 
bipyramid.

Such distorted decahedral structures have previously been seen for 
a number of metal 
potentials,\cite{Doye98c,Lloyd98,Lloyd00,Wilson00,Soler01,Darby02,Lai02,Sebetci03} 
particularly those that have a tendency to disorder.
Indeed, Soler {\it et al.} have identified such structures as
particularly important for amorphous gold clusters, and
have used them as a basis to suggest schemes to produce 
potentially new magic number clusters for larger sizes.\cite{Soler01}

As the size of the clusters increase, so of course must the number of atoms in
the interior. Zn$_{25}$ is typical of most of the clusters with
two internal atoms. It can be considered as two interlocking 
distorted decahedra. In the view shown in Fig.\ \ref{fig:Znselect}
a Zn$_{18}$-like fragment can be clearly seen. The exception to these
type of structures occurs at Zn$_{28}$. The six-fold symmetric structure
is a continuation of Zn$_{14}$ and has four stacked hexagonal rings 
arranged in an antiprismatic fashion. One would expect such a structure to
have three internal atoms, but instead there is a vacancy in the very centre 
of the structure. Again, this is to avoid internal atoms with values
of $\bar\rho$ that are too large. 

Zn$_{30}$ and Zn$_{34}$ are typical of structures with three and
four internal atoms. The overall shape reflects the triangular 
and tetrahedral arrangement of the internal atoms, but it is 
again hard to detect any overall order.
Zn$_{36}$ provides an interesting exception. The surface atoms have
basically the same geometry as a recently identified 38-atom
polytetrahedral structure.\cite{Doye01a,Doye03b} 
However, there are only four (not six) internal atoms, and these are
arranged as a planar rhombus.

First at $N$=35 and then from $N$=38 a new series of distorted decahedral structures begins.
Again they are based on a Marks decahedra that has been created from
a pentagonal bipyramid (in this case with 54 atoms) by introducing 
reentrant grooves that are one 
layer deep. A similar mechanism of distortion is again seen. 
Atoms that are on the equatorial edges of the cluster 
either side of a groove are drawn closer together, although in this case a contact is not
actually formed---the distortion is a diamond-square rather than
a diamond-square-diamond process. 
The clusters at $N$=35, 39, 41, 45, 47 and 49 in Fig.\ \ref{fig:Znselect}
provide examples of structures in this sequence. 
In most of the examples, two such distortions are present at adjacent 
grooves, and so the whole edge between these grooves is brought closer
to the centre of the cluster, forming new contacts with the closest interior atoms. 
As can be seen for the second view of the complete 49-atom Marks decahedron, this 
lead to a splaying of the planes of pentagonal rings that would 
otherwise have been parallel.

For these larger clusters, these distortions represent a smaller
perturbation on the overall structure than for $N$=16-21. Indeed, they
are barely visible when the cluster is viewed down the quasi-fivefold axis.
All that can be seen is a slight opening of the angle at the 
reentrant groove.

Comparing the energetics of the distorted and undistorted 49-atom Marks decahedron
shows that the main improvement arising from the distortion is for the interior atoms.
The movement of the equatorial edges involved in the distortion closer to the centre of the cluster
compresses the rest of the surface somewhat and increases the coordination number
of some of the interior atoms. This leads to a structure where the ratio of the 
nearest neighbour distances for the interior atoms to those of the surface atoms
increases from 1.049 to 1.060. Thus, there is less need for the surface to shrink
inwards and compress the core in order to improve the $\bar\rho$ values of the surface atoms.
Consequently, $\bar\rho_{\rm bulk}$, the average $\bar\rho$ value for the interior atoms, 
decreases from 9.197 to 8.756, i.e.\ closer to $\bar\rho_{\rm xtal}$, and so the energy of 
the interior atoms decreases.
By contrast the distortion has little overall effect on the energy of the surface atoms. 
Generally, the atoms in the equatorial plane, especially those close to the distortion, improve
their energies, but many of the other surface atoms lose out due to the breaking or stretching
of contacts in the axial direction. 

This decahedral series of structures is interrupted at N=42 and 43 by 
structures with four-fold symmetry. These clusters are loosely related to
the 44-atom fcc octahedron, but with one or two opposite vertices removed.
The top half of the cluster is then twisted with respect to the bottom to
give a slightly buckled outer layer.

Beyond $N$=50 these groove-bridged Marks decahedra are no longer most stable. 
Instead, there is a size range where the structures typically have no overall
order, but where motifs that resemble fragments of Marks decahedra and 
Mackay icosahedra are evident. Zn$_{54}$ provides a typical example, 
whereas Zn$_{52}$ is a structure with more order evident. From one
side Zn$_{52}$ has perfect Mackay icosahedral order. What has occurred to 
the other side can be determined by comparing to the 50-atom incomplete Mackay 
icosahedron shown in Fig.\ \ref{fig:ref_struct}.
The two contacts that complete the five-fold rings around the two
empty vertices of the Mackay icosahedron have been broken, removing
some of the tension in the outer layer of the Mackay icosahedron that
makes it so energetically unfavourable.\cite{hiMackay} Then two atoms (oriented horizontally
with respect to the views in the figures) are inserted into the
coordination shell of the central atom, increasing its coordination number.
Such a structure has previously been located for Au$_{52}$ modelled 
by a Sutton-Chen potential.\cite{Doye98c}

From the particularly stable structures based on the 18-atom and 
49-atom Marks decahedra, it is not surprising that there is another series
of magic numbers leading to the 100-atom Marks decahedron. Again this Marks 
decahedron is formed from a pentagonal bipyramid by the introduction
of grooves of depth one layer.
More generally, one can use this trend to predict potential magic number
clusters outside the size range of this study.
The sizes for such complete Marks decahedra are given by
\begin{equation}
N={5\over 6} n^3 + 5n^2 + {61\over 6} n +2 
\end{equation}
where $n$ is the number of atoms on an equatorial edge of the Marks decahedron.
This gives $N$=18, 49, 100, 176, 282, $\ldots$

The first of this set of groove-bridged decahedral structures occurs at 
$N$=75 and the last at $N$=101. Representative examples at 
$N$=75, 77, 79, 83, 86--88, 92, 96 and 100 are shown in Fig.\ \ref{fig:Znselect}. 
At the smallest sizes the decahedra are still quite asymmetric.
From Fig.\ \ref{fig:EvN} one can see that the particularly stable sizes 
occur at $N=100-4m$, where $m$=0--3. These structures can be formed from Zn$_{100}$ by the sequential 
removal of 4-atoms from each groove to give structures that have $m$ grooves that are two
layers deep. It is interesting to know that the complete Marks decahedron is not in fact 
the most stable of these structures, but that a slight asymmetry is preferred.

Comparing the energetics of the 100-atom distorted and undistorted Marks decahedra reveals
a similar story to Zn$_{49}$. The main stabilization of the distorted structure 
is due to the decrease in $\bar\rho_{\rm bulk}$ from 8.875 to 8.789.
It is also interesting to analyse the reasons for the stability of structures 
at $N$=88 and 92 with four-coordinate surface atoms. 
Firstly, this is because a four-coordinate atom can compensate for its low coordination
by having very short nearest-neighbour distances, thus achieving a reasonable $\rho$ value;
e.g.\ 7.519 for Zn$_{92}$. Secondly, this additional short contact reduces the need for the
four surface atoms in contact with the adatom to shrink inward and compress the adjacent
interior atoms. For Zn$_{92}$ the $\bar\rho$ values for these atoms only increase by 0.313,
despite this extra contact, because the pair separations for the other six contacts 
have expanded by 2.5\%.

\begin{figure*}
\includegraphics[width=15cm]{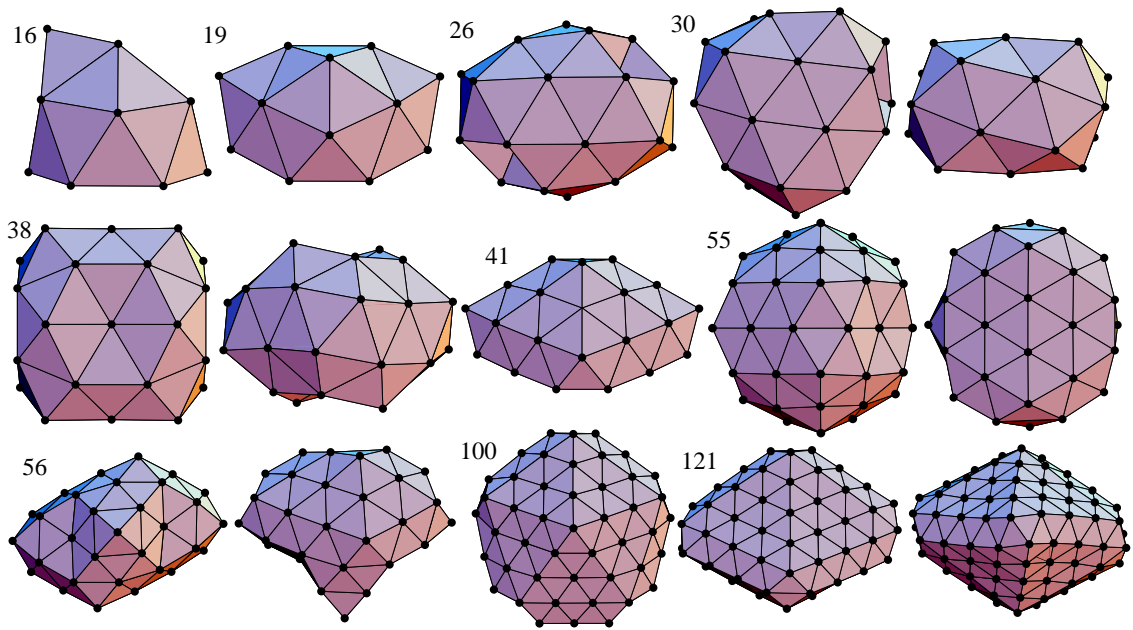}
\caption{\label{fig:Cdselect}A selection of the putative global minima for Cd$_N$ that 
differ from those for Zn$_N$.}
\end{figure*}

The structures at $N$=63-70 are also closely based on the 100-atom Marks
decahedron, but with a further distortion. From the first view of Zn$_{64}$ 
one can see the resemblance to the 63-atom structure in Fig.\ 
\ref{fig:ref_struct} that is an asymmetric fragment of the 100-atom 
distorted Marks decahedron. The second view of Zn$_{64}$ shows the 2-fold axis. 
From the equivalent view of the 63-atom structure, it is clear that 
an extra atom has been added to the column of three capping 
atoms (at the bottom of the first view and in a vertical line in the centre of the 
top half of the cluster in the second view), 
and then a small twist has been given to the two halves of 
the structure. Zn$_{69}$ has basically the same structure but with an increase 
in the size of one of the decahedral faces.

Based on the favourability of small distorted Marks decahedra (similarly to 
those we see at $N$=16--21) for gold clusters 
modelled by a Gupta potential, Soler {\it et al.}  
suggested a means for using this structural pattern to generate
potentially stable large clusters.\cite{Soler01}
A 55-atom Mackay icosahedron can be generated from a 13-atom icosahedron by
the addition of atoms at the centre of each nearest-neighbour contact.
By the same process, an 80-atom structure can be generated from an 18-atom 
distorted Marks decahedron with one of the grooves bridged.
Soler {\it et al.} envisaged that this structure would be most 
stable when the three four-coordinate atoms in this 80-atom structure were 
removed.\cite{Soler01}
However, the suggested structure is actually the global minimum for Zn$_{80}$, 
reflecting the stabilization of low-coordinate atoms 
that is possible for this potential.

An alternative way to generate the Zn$_{80}$ structure is to introduce reentrant grooves 
two layers deep into the 100-atom pentagonal bipyramid. This Marks decahedron
is then distorted to produce four new contacts that bridge the two capping 
square pyramids adjacent to a groove. This distortion is analogous to that for 
bridging a groove one layer deep.
In the region of the distortion, the structure looks locally like 
a 55-atom Mackay icosahedron.

Bridging a groove that is two layers deep is generally less favourable 
than bridging one that is one layer deep.
There are only five examples for $N\le$100. The others occur for Zn$_{71-73}$ and Zn$_{84}$.
Zn$_{71}$ and Zn$_{84}$ are quite interesting because they involve distortions of
grooves that are both one and two layers deep.
For example, Zn$_{84}$ can be formed from Zn$_{80}$, by first filling in one of the grooves, so that
it is then only one layer deep and then applying the distortion to this groove.

After the completion of the previous decahedral series associated with the Marks decahedra
at $N$=18 and 49, there are size ranges where the majority of clusters have no
discernible overall order at $N$=22--34 and $N$=51--62, before structures
based on the next Marks decahedron become lowest in energy. 
However, even though the next complete Marks decahedron is at $N$=176, beyond $N$=101
structures based on this larger Marks decahedron are immediately lowest in energy.
From $N$=102--117 (excepting $N$=108) the zinc clusters are
asymmetric groove-bridged decahedra with the same twist distortion as that for $N$=63-70. 
Like for Zn$_{64}$ this leads to structures with a two-fold axis of symmetry at 
$N$=111 and 115 (see the second views of these structures in Fig.\ \ref{fig:ref_struct}). 

The first global minima that are fragments of the 176-atom Marks decahedron with only 
distortions bridging grooves one layer deep occur at $N$=122 and 123. 
There are also a number of structures, where grooves two layers deep are bridged; e.g. 
$N$=108, 120, 121 and 125. The structures at $N$=108, 121 and 125 are slightly more 
complicated, because the reentrant $\{111\}$ faces in the grooves of the undistorted 
Marks decahedra are not triangular but instead have two atoms along the outer apical edge. 
This is illustrated for Zn$_{108}$, which is based on the 111-atom Marks decahedron shown in
Fig.\ \ref{fig:ref_struct}. On bridging this groove, it is favourable to remove three atoms
that lie on the $\sigma_h$ mirror plane of the Marks decahedron, and to twist the top and bottom 
of the cluster slightly to remove a small $\{100\}$-type face that would otherwise result.

As has already been noted, for just over half the cadmium clusters in the size range considered, 
the global minimum have the same structures as the zinc clusters. 
A selection of examples where the global minimum differs from zinc 
are depicted in Fig.\ \ref{fig:Cdselect}.
The global minima are again dominated by distorted Marks decahedra, 
although the positions of capping atoms (e.g.\ Cd$_{16}$ and Cd$_{19}$) 
or the number of grooves that are bridged (e.g.\ Cd$_{100}$) may be different.
To take Cd$_{100}$ as an example, this structure is second lowest in energy for zinc, 
because, although it has a better embedding energy than the global minimum, this is more than offset
by a worse pair energy. However, because of the reduced magnitude of the pair
interactions for cadmium this structure is now lowest in energy.

Another difference between the two systems is the more pronounced stability 
of the twisted Marks decahedra for cadmium,
as indicated by Fig.\ \ref{fig:EvN} and the larger size ranges ($N$=63--72 and 102--121)
for which these structures are most stable. A further difference is that decahedral structures
where grooves of depth two are bridged are less common---there are only three examples
for cadmium clusters in the size range considered here.

The least coincidence between the zinc and cadmium global minima probably occurs in the 
size windows between the series of distorted Marks decahedra where the structures often have no
overall order. This probably reflects the large number of disordered structures that 
only have small differences in energy. Examples of clusters from these size ranges with 
$N$=26, 30, 38, 55 and 56 are shown in Fig.\ \ref{fig:Cdselect}.

\section{\label{sect:conc}Conclusion}

I have analysed the structures of the global minima up to $N$=125 for two metallic 
potentials that have been previously found to have a particularly strong tendency to 
exhibit disordered clusters. This study confirms that the clusters
exhibit none of usual ordered forms for materials that are close-packed in bulk.
Instead, the majority of the clusters are based on distorted oblate Marks
decahedra, but where the distortions are well-defined. There are series of structures
associated with the 18-, 49-, 100- and 176-atom complete Marks decahedra. However, in 
between there are size windows where the majority of the clusters have no overall order,
except for $N>100$ where there is a direct transition between structures based on the 100- and
176-atom Marks decahedra.

Because the effective pair potentials of these zinc and cadmium potentials are very 
shallow, thus providing relatively little constraint on the pair separations, there
is a strong tendency for the surface atoms to contract inwards in order for these atoms 
to obtain a better many-body embedding energy. The resulting compression of the interior of the 
cluster causes the conventional structures to be disfavoured,\cite{Soler00} 
because it results in a $\bar\rho_{\rm bulk}$ that is significantly larger than
$\bar\rho_{\rm xtal}$. Instead structures for which the nearest-neighbour distances for the
surface atoms are naturally longer than those for the interior atoms are likely to
be favoured, because this allows the difference between $\bar\rho_{\rm bulk}$ 
and $\bar\rho_{\rm surf}$ to be reduced, bringing them both closer to the optimal value.
The distortions of the oblate Marks decahedra help to achieve just this.

This paper should be seen in the context of a growing research program
that has sought to understand how the observed structures for metal clusters depend 
on the form of the many-body potential.\cite{Soler00,Baletto02b,Hendy02,Doye03a,Doye03d}
Even for simple potentials of the embedded-atom (or glue) form (Eq.\ \ref{eq:EAM}) this leads to 
significant complexity. The current results further 
highlight how many-body potentials can lead to the stabilization of unusual 
structural forms.\cite{Soler00,Soler01,Hendy02,Doye03a,Doye03d}
Of course, at sufficiently large sizes the bulk structure, in this case hexagonal close-packed, 
should become most stable, but the energy differences found in this study suggest 
that the current clusters are far from this limit.

The emphasis in this paper has been on the zinc and cadmium clusters as model systems with
a strong tendency to disorder. As noted in Sect.\ \ref{sect:potential},  with 
the Gupta potential it is difficult to capture the large $c/a$ ratio for 
Zn and Cd without introducing discrepancies for other properties.
I have also performed a brief set of global optimization runs using 
Cleri and Rosato's alternative Gupta potential for Cd, where the $c/a$ ratio 
was allowed to vary during the fitting. For the selection of sizes
I tested the usual ordered forms, i.e.\ close-packed, icosahedral and decahedral, 
were always most stable. 
Reyes-Nava {\it et al.} also found these cadmium clusters to have strong
first-order-like melting transitions, again reflecting the strong ordering
for this alternative potential.\cite{ReyesNava03}
Given this strong dependence on the potential parameterization, 
one should be somewhat sceptical about the abilities of 
these potentials to provide realistic models for zinc and cadmium clusters.
However, the tendency to disorder for these clusters
seems to be reproduced by electronic structure calculations using 
density-functional theory.\cite{Michaelian02}

There has been little other relevant work. 
Experiments on these clusters have focussed on their electronic shell structure, 
\cite{Katakuse86,Ruppel92} rather than their geometries,
and their are only a few theoretical studies to go beyond very small sizes.
For cadmium clusters electronic structure calculations have been performed 
up to 20 atoms.\cite{Zhao01} 
These found Cd$_{13}$ to be an icosahedron and the larger clusters to 
be based upon this structure.\cite{Zhao01} 
Ramprasad and Hoagland studied zinc clusters using a many-body potential 
of the embedded-atom form (Eq.\ \ref{eq:EAM}), 
but where the effective interactions looked very different from 
the current Gupta potential.\cite{Ramprasad93}
Of the series of candidate structures that they reoptimized, they also found icosahedral structures 
to be lowest in energy.

\begin{acknowledgments}
The author is grateful to the Royal Society for the award of a University Research Fellowship.
\end{acknowledgments}


\begin{thebibliography}{52}
\bibitem[{als()}]{alsotruncate}
\bibinfo{note}{In many cases, although not for the presents clusters, to
  produce particularly stable Marks decahedra, as well as the introduction of
  grooves into the pentagonal bipyramids, it is also important to truncate the
  equatorial edges to reveal $\{100\}$ facets.}
\bibitem[{hiM()}]{hiMackay}
\bibinfo{note}{The 55-atom Mackay icosahedron is in fact $0.828$\,eV higher in
  energy than the Zn$_{55}$ global minimum.}
\end{thebibliography}
\end{document}